\documentclass[prb,twocolumn,showpacs]{revtex4}
\usepackage{graphicx}
\usepackage{amsmath}
\usepackage{amssymb}
\usepackage{epstopdf}

\begin{document}
\title{Identifying Quantum Topological Phases Through Statistical Correlation}

\author{Hao Wang$^{1}$, B. Bauer$^{2}$, M. Troyer$^{2}$, V. W. Scarola$^{1}$}
\affiliation {
$^{1}$Physics Department, Virginia Tech, Blacksburg, Virginia 24061, USA\\
$^{2}$Theoretische Physik, ETH Zurich, 8093 Zurich, Switzerland
}

\begin{abstract}
We theoretically examine the use of a statistical distance measure,
the indistinguishability, as a generic tool for the identification
of topological order.  We apply this measure to the toric code and
two fractional quantum Hall models.  We find that topologically ordered
states can be identified with the indistinguishability for both
models.  Calculations with the indistinguishability also underscore
a key distinction between symmetries that underly topological order
in the toric code and quantum Hall models.
\end{abstract}

\pacs{03.65.Vf, 73.43.-f, 05.30.Pr}
%03.65.Vf   Phases: geometric; dynamic or topological
%73.43.-f   Quantum Hall effects
%05.30.Pr   Fractional statistics systems (anyons, etc.)
\maketitle

\section{Introduction}

Conventional types of quantum order can be characterized by local
symmetries. Topological quantum order, in contrast, defies
characterization by local operators \cite{wen:1990a}.  Topological
order owes its structure to non-local properties and
therefore depends on the surface on which it is placed.  Examples of
topologically ordered quantum states include the ground state
of Kitaev's toric code model \cite{kitaev:2003} and quantum Hall
states \cite{topofqhe}.  Such states are not characterized by
simple, local order parameters. Analyses based on system entanglement entropy and other non-local properties have been used to study these states.\cite{tsomokos:2009,haque:2007}

Wavefunctions (e.g., the Laughlin state \cite{laughlin:1983})
capture the essential properties of some of the fractional quantum
Hall (FQH) states. Indeed, full microscopic analyses are typically
done with wavefunctions in efforts to accurately capture the low
energy physics of insoluble quantum Hall
models \cite{dassarma:1997,jain:2007}.  These wavefunctions, in turn,
describe incompressible quantum liquids with no simple local order
parameter.

Kitaev has constructed exactly soluble spin models, the toric code
\cite{kitaev:2003} and honeycomb\cite{kitaev:2006} models, to
analytically probe topological order.  These two-dimensional models
exhibit one-dimensional string symmetries that underly topological
order.  A comparison between symmetries in these spin models and
certain symmetries of the FQH regime \cite{haldane:1983} has been
recently drawn \cite{nussinov:2009}.  We ask if one can use
numerical methods to generically identify and compare topological
order in both types of models.

The indistinguishability
\cite{helstrom:1976,fuchs:1999,korsbakken:2007} was recently
proposed \cite{bauer:2010} as a tool to probe complex quantum
states.  The indistinguishability is a statistical distance measure
that yields the probability of making an error in an $n$-particle
measurement in an attempt to distinguish two states.
Ref.~\onlinecite{bauer:2010} used explicit calculations on
one-dimensional spin models to test if this measure can act as an
effective non-local order parameter to identify quantum states.
Scaling relations were found in transitions between states in the
quantum Ising model and the bilinear-biquadratic Heisenberg chain
without making recourse to local order parameters.  Phases and phase
transitions were instead identified using ansatz states.

In this paper we study the indistinguishability as a method to
identify topological quantum order in two-dimensional models.  We
study the toric code and models of the FQH regime. We find that in
the toric code the indistinguishability reveals distinct topological
sectors  and the one-dimensional nature of the symmetries defining
each sector.   We then use the indistinguishability to underscore a
key difference between topological order in the toric code and the
FQH regime.  By diagonalizing models of the FQH effect we show that
distinct topological sectors (and distinct FQH states in general)
differ in that symmetry operators must span the \emph{entire} system
rather than just one-dimensional operators.  The measure can be used
to identify mechanisms of topological ordering in more non-trivial
models where symmetries and a complete characterization of states
have not been performed.

In Section~\ref{distinguishability} we review the
indistinguishability as a measure of distinct quantum orders. In
Section~\ref{toriccode} we examine the scaling behavior of the
indistinguishability in the toric code.  In Section~\ref{fqhe} we
examine the scaling of the indistinguishability in FQH models of the
Laughlin, charge density wave \cite{koulakov:1996}(CDW), and Moore-Read\cite{moore:1991} states.  We summarize in
Section~\ref{summary} with a comparison of results for both sets of
models.

\section{Indistinguishability}
\label{distinguishability}

The indistinguishability is based on a quantum information measure
of quantum state distinguishability
\cite{helstrom:1976,fuchs:1999,korsbakken:2007}.  We define the
indistinguishability $I_{n}(\text{A:B})$ of two $N$-particle states,
$\Psi_\text{A}$ and $\Psi_\text{B}$, as the probability
of making an error in distinguishing the two states with an $n$-particle
measurement:
\begin{eqnarray}
I_{n}(\text{A:B})=\frac{1}{2}-\frac{1}{4}\text{Tr} \vert
\rho^{(n)}_{\text{B}}-\rho^{(n)}_{\text{A}} \vert, \label{In}
\end{eqnarray}
where $\text{Tr} \vert \Omega \vert$ is the trace norm of
$\Omega$ and $\rho^{(n)}=\text{Tr}_{N-n}\left(\rho\right)$ is
the $n$-particle reduced density matrix and $\text{Tr}_{N-n}$ denotes the partial trace over $n$
particles. Interpreting the density
matrix as a probability distribution, the last term in Eq.~\ref{In} can
be identified with a well-known statistical distance measure, the
Kolmogorov distance. When
$I_{n}$ is zero, two states are distinguishable and the
ansatz state $\Psi_\text{A}$ is a poor approximation
to $\Psi_\text{B}$.  However, when it is non-zero, there is a finite
probability that an $n$-particle measurement can not distinguish the
two states.  $I_{n}=1/2$ corresponds to the maximum indistinguishability, implying two identical
states with unitary wavefunction overlap when $n=N$.  In contrast to the entanglement entropy used in the FQHE regime, here the state indistinguishability yields a single number that
quantifies the ability of an optimally chosen set of $n$-particle
correlators to distinguish two states \cite{helstrom:1976,
fuchs:1999}.  $1-I_{n}$ gives the probability that an optimally
chosen correlation function involving at most $n$ particles will be
able to distinguish the  two states.

We use $I_n$ to quantify the degree of \emph{indistinguishability}
of two states imposed by underlying correlators in an $N$-particle system. In cases where
a small constant value of $n \sim \mathcal{O}(1)$ suffices to characterize the correlators (i.e.,
two states can be distinguished locally), we
define $I_n$ to be \emph{intensive} in $N$. Such two states belong to the same
$n$-particle \emph{correlator class} if $I_n$ remains finite in the thermodynamic
limit (i.e., as $N \rightarrow \infty$).
On the other hand, if two states can not be distinguished locally and therefore
$n$ needs to scale with $N$, we define $I_n$ to be \emph{extensive}.
In this situation, we use the scaling of $n$ with $N$ to identify
correlator classes \cite{bauer:2010}.  The precise scaling behavior of
$n$ with $N$ (e.g., $n \sim \mathcal{O}(N)$ or
$n \sim \mathcal{O}(\sqrt{N})$) provides us with a key feature to reliably distinguish phases.

In the following, we explore the scaling of the indistinguishability
between topologically ordered quantum states.

\section{Indistinguishability in the Toric Code}
\label{toriccode}

\subsection{Review of the Toric Code}

\begin{figure}
\centerline{\includegraphics [width=1.5 in] {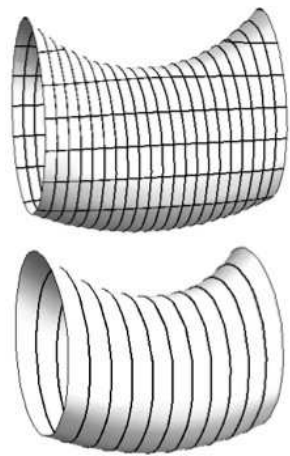}} \caption{ Top:
Section of torus depicting the two-dimensional basis of the toric
code in real space. The sites sit on bonds between vertices to form
a square lattice. Bottom:  Section of torus depicting
two-dimensional basis of a single Landau level in real space.  Basis
states form a periodic array of rings in the Landau gauge. }
\label{compare}
\end{figure}

The toric code Hamiltonian was constructed as an exactly soluble
model with a topologically ordered ground state and anyonic excitations
\cite{kitaev:2003,castelnovo:2010}.  We briefly review the model and discuss its
symmetry properties.  The model is given by:
\begin{equation}
H_{T}=-\sum_{v}\prod_{j\in v}\sigma_{j}^{x}-\sum_{p}\prod_{j\in
p}\sigma_{j}^{z},
\end{equation}
where {\boldmath$\sigma$}$_j$ denotes Pauli matrices at sites $j$ on
bonds of the square lattice.  The first product is over the four
sites surrounding the vertex $v$ while the second product is over
the four sites around each plaquette $p$ (Figure~\ref{compare}).

When placed on a torus, the model possesses two distinct
one-dimensional $\mathbb{Z}_2$ symmetries. The operators
$\prod_{j\in w'}\sigma_{j}^{x} $ and $ \prod_{j\in w}\sigma_{j}^{z}
$ both commute with $H_{T}$ where $w'$ is a loop along vertices and
$w$ is a loop along bonds.  These one-dimensional operators form
closed loops around either cycle of a torus.  They can be used to
classify topological ground state sectors.

The ground state of the toric code is then given as the equal-amplitude
superposition of vortex-free states:
\begin{equation}
|\Psi_i \rangle = \sum_{|\xi\rangle \in \chi_i} f_i |\xi\rangle,
\label{psitoric}
\end{equation}
where $\chi_i$ are four spaces of such vortex-free configurations
distinguished by the expectation value of the operator:
$ \prod_{i \in w} \sigma_i^z$,
for two in-equivalent non-contractible loops $w_1, w_2$ wrapping
around the torus in two different directions.  $f_i$
is a normalization factor which is equal for all sectors. By a
vortex-free configuration, we mean a basis state $|\xi\rangle$ for
which $\prod_{i \in \delta p} \sigma_i^z = +1$ for all plaquettes
$\delta p$.

The toric code exhibits a phase transition under a magnetic field.
This perturbation breaks the one-dimensional $\mathbb{Z}_2$
symmetries and can destroy topological order if it is strong enough.
Numerical studies of ground state degeneracies and other indirect
measures of topological order show a robust phase transition
\cite{trebst:2006} from the topologically ordered phase to a
classically ordered phase with increasing magnetic field.  A more
recent study used a topological fidelity measure to observe the same
transition by extracting finite size scaling information related to
the one-dimensional $\mathbb{Z}_2$ symmetries
\cite{tagliacozzo:2010}.

\subsection{Computed Indistinguishability}

Given the above ground states of the toric code, we can analytically
compute the indistinguishability between two topologically distinct
sectors. We consider a square lattice $\mathcal{L}$ with spins
located on $L$ bonds along each dimension and $N = 2 L^2$ sites. A
block $\mathcal{Q}$ of $n$ sites is chosen for calculating $I_n$.
The remaining sites in the lattice are denoted as $\mathcal{R}$,
i.e., $\mathcal{L} = \mathcal{Q} \cup \mathcal{R}$.

To compute $I_{n}$ we must find $\varrho_{A}^\mathcal{Q}$ and
$\varrho_{B}^\mathcal{Q}$, the reduced density matrices on a subset
$\mathcal{Q} \subset \mathcal{L}$ for two different states $A$ and
$B$, respectively.  From Eq.~\ref{psitoric} we find that the matrix
elements of $\varrho^\mathcal{Q}_A - \varrho^\mathcal{Q}_B$ are
given by:
\begin{align}
& \langle v_\mathcal{Q} | \varrho^\mathcal{Q}_A - \varrho^\mathcal{Q}_B | w_\mathcal{Q} \rangle  \nonumber \\
= \sum_{|u_\mathcal{R}\rangle} f^2 \Bigl\{ &\sum_{|\xi_1\rangle,|\xi_2 \rangle \in \chi_A} \langle v_\mathcal{Q} u_\mathcal{R} | \xi_1 \rangle \langle \xi_2 | w_\mathcal{Q} u_\mathcal{R} \rangle \nonumber\\
 - &\sum_{|\xi_1\rangle,|\xi_2 \rangle \in \chi_B} \langle v_\mathcal{Q} u_\mathcal{R} | \xi_1 \rangle \langle \xi_2 | w_\mathcal{Q} u_\mathcal{R} \rangle\Bigr\} \nonumber \\
= \sum_{|u_\mathcal{R}\rangle} f^2 \Bigl\{ & \delta_A(| v_\mathcal{Q} u_\mathcal{R} \rangle) \delta_A( | w_\mathcal{Q} u_\mathcal{R} \rangle ) \nonumber \\
 - &\delta_B( | v_\mathcal{Q} u_\mathcal{R} \rangle) \delta_B( |w_\mathcal{Q} u_\mathcal{R} \rangle) \Bigr\}. \label{toricIn}
\end{align}
Here, the states $|u_\mathcal{R}\rangle$ are all basis states on the
sublattice $\mathcal{R}$, and $\delta_A (|\phi\rangle) = 1$ if
$|\phi\rangle \in \text{span}(\chi_A)$, 0 otherwise.

The above expression shows that if $\mathcal{Q}$ supports two
in-equivalent loops $w_1, w_2$, all sectors can be distinguished, as
expected. If it only supports one such loop, only half of the
sectors can be distinguished. If it does not wrap around the
boundary, no sectors can be distinguished. $I_n$ is always either 0
or 1/2.  The above explicit calculation therefore shows that for a
wisely chosen $\mathcal{Q}$, such that it wraps the boundary ($w_1,
w_2 \in \mathcal{Q}$), correlators of size $n =
\mathcal{O}(\sqrt{N})$ are sufficient to reliably distinguish
topological sectors.

We now ask how many measurements on \emph{randomly} chosen spins are
needed to distinguish topological sectors of the toric code. For
simplicity, we consider only the case of distinguishing two sectors,
i.e., we look for clusters wrapping around the torus in one
non-trivial way. We seek the probability $\Pi(p)$ that a fraction
$p$ of randomly chosen sites forms a cluster that wraps around the
boundary.  This is the problem of percolation with periodic boundary
conditions. For this problem, it is well-known that a critical $p_c$
exists such that in the thermodynamic limit, $\Pi = 1$ for $p > p_c$
and $\Pi = 0$ otherwise. The critical behavior is in fact identical
to that of standard percolation with free boundary
conditions~\cite{stauffer1994,watanabe1995}. These well-known
results from percolation theory indicate that in order to
distinguish sectors of the ground state based on purely randomly
chosen sites, a cluster size $n \sim \mathcal{O}(N)$ is necessary.

A different situation occurs if we choose sites randomly, but as a contiguous
blocks. The probability for a contiguous cluster of size $n$ to wrap around the
boundary, which we denote as $\varrho$, is given by:
\begin{equation}
\varrho(n) = \int_{p_c}^1 dp\ \delta(NP(p)-n),
\end{equation}
where $P$ is the probability for one site to lie in the percolating
cluster for a completely random choice of sites. We then have $I_n =
(1-\varrho(n))/2$. We do not expect a sharp transition to appear in
this quantity because there is a finite but exponentially small
probability for a random block of size $n \geq \sqrt{N}$ to wrap
around the boundary.

Scaling theory dictates that the behavior of $P$ in the
thermodynamic limit and in the critical region is governed by $P
\sim (p-p_c)^\beta$. The divergence of the correlation length is
described by $\xi \sim (p-p_c)^{-\nu}$; however, on finite systems
this is bounded by $L$ and therefore $(p-p_c) \sim L^{-1/\nu}$. We
then have $P \sim L^{-\beta/\nu}$ or, equivalently, a critical
cluster size $n_c \sim L^{2-\beta\nu} = L^D$, where $D$ is the
fractal dimension. If one were to grow only one cluster in the
system, the probability for this cluster to percolate should
increase rapidly at $n \sim \mathcal{O}(L^D)$. In two dimensions,
the value of $D$ is $91/48$. We can therefore expect that a
contiguous cluster of size
\begin{equation}
n \sim \mathcal{O}(L^{91/48})
\end{equation}
is sufficient to distinguish two sectors of the ground state.

\begin{figure}[t]
\centerline{\includegraphics [width=3.2 in] {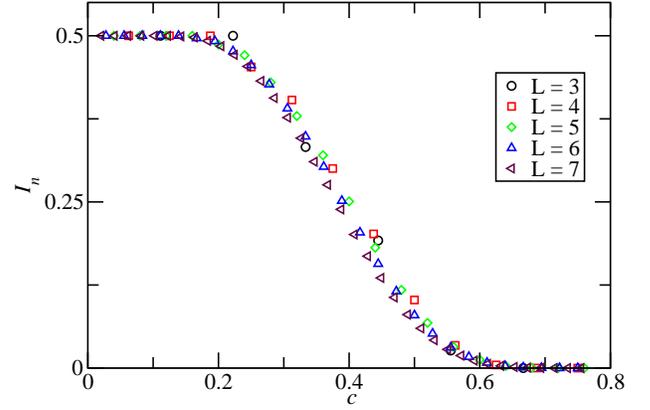}} \caption{(Color
online) Plot of the $n$-particle indistinguishability versus $c=n/N$
for several different system sizes computed using Monte Carlo
selection of random but contiguous collections of spins for the
toric code on a two-dimensional periodic lattice with $N=2L^{2}$
spins.  The graph shows data collapse and a linear scaling of $n$
with system size, $N$, in contrast to a $N^{1/2}$ scaling for
properly chosen spins (Eq.~\ref{toricIn}). } \label{fig_loop}
\end{figure}

To verify the above statement we compute $I_{n}$ explicitly using a
direct-sampling Monte Carlo method. We draw the configurations of a
cluster with $n$ connected sites from a uniform distribution and
measure the probability for such a cluster to support a loop
wrapping around the boundary, $P(\text{loop}) = \varrho(n)$. The
results of $I_n$ versus linear scaling ratio:
\begin{equation}
 c=\frac{n}{N},
 \end{equation}
 for several $L$ are shown in Figure \ref{fig_loop}.
The data collapse beyond a regime where finite-size effects are
relevant, which agrees with the expected scaling $n \sim
\mathcal{O}(L^D)$. The difference between $L^{91/48}$ and $L^2$ is
too small to be distinguished numerically.

We have thus shown that the indistinguishability reveals the size of
the operators required to identify topological sectors.  For suitably
chosen blocks we find $n = \mathcal{O}(\sqrt{N})$ whereas randomly
chosen sites lead to $n = \mathcal{O}(N)$. In the case where a
random, but contiguous choice of sites is made, the necessary block
size is $n = \mathcal{O}(L^{91/48})$. $I_{n}$ thus yields
topologically relevant information without requiring a precise
identification of the non-local symmetries defining each sector. We
now turn to models of the FQH effect that, in some limits, do not
have exact solutions.

\section{Indistinguishability in the Fractional Quantum Hall Regime}
\label{fqhe}

We consider a two-dimensional electron gas (2DEG) on the surface of
a torus under a magnetic field perpendicular to the surface. In a
strong magnetic field, electrons occupy highly-degenerate and
energetically distinct Landau levels (LLs). At a fractional LL
filling, $\nu$, ideal interaction models can generate topologically
ordered ground states without defining local symmetries. Two
examples include the Abelian Laughlin states \cite{laughlin:1983} at
$\nu=1/3$ from short range pair interactions \cite{haldane:1985} and
the non-Abelian Moore-Read state \cite{moore:1991}  at $\nu=5/2$
from short range three-body interactions
\cite{greiter:1991,rezayi:2000}.

In the torus geometry, these ground states are degenerate in
multiple folds and show a finite energy gap from excited states,
thereby suggesting topologically ordered states.  A key question
then arises.  Is there a simple correlation function (e.g., a
one-dimensional chain operator with $n\sim \sqrt{N}$ as for the
toric code) that defines the topological sectors in the quantum Hall
regime?  A well-known result by Haldane \cite{haldane:1983}
discovered just such a symmetry.  A product of translation operators
around one toric cycle indeed connects distinct topological sectors.
(For a review of this work and its connection to topological sectors
see Ref.~\onlinecite{stone:1992}.)  This center of mass operator
requires all particles for its construction, in contrast to the
chain operators identified in the toric code.

Peculiarities of the lowest LL basis require that a symmetry
spanning at least one dimension must incorporate all particles.  To
see this we consider basis states on a section of the torus as shown
at the bottom of Figure~\ref{compare}. The Landau gauge basis states
form periodic rings around the torus.  Operators constructed from
one-dimensional translations of these rings will encompass the
entire system.  One can show that there are no lowest LL basis
states that are both orthogonal and localized in two dimensions
\cite{zak:1997}.  As a result, apparent one-dimensional symmetries
must span all particles in two-dimensional lowest LL
systems\cite{nussinov:2009}.

The apparent lack of true one-dimensional symmetries suggests that
\emph{all} FQH states are best characterized by $n \sim N$
correlation functions, i.e., wavefunctions, in systems without
edges.  We verify this assertion using the indistinguishability to
compare a variety of different FQH states on the torus. We compare
these states by first constructing generator models, diagonalizing
these models, and then numerically computing $I_{n}$.

\subsection{Modeling Fractional Quantum Hall States}

We now review the Coulomb model of  the FQH regime and ideal models
that generate FQH states. Periodic boundary conditions for magnetic
translational operators are imposed with a quantized flux $N_{\phi}$
through the unit cell. The magnetic length $\ell$ is taken as the
unit length and the energy is in Coulomb units, $e^2/4\pi \epsilon
\ell$. In the absence of LL mixing, the Hamiltonian
of a 2DEG of $N$ particles interacting through the Coulomb
interaction can be projected into the topmost LL with the filling factor
$\tilde{\nu}=N/N_{\phi}$\cite{rezayi:2000}:
\begin{eqnarray}
  H_{c}=
    \frac{2}{N_{\phi}} \sum_{i<j}\sum_{\mathbf{q}} e^{-q^2/2} e^{i \mathbf{q} \cdot (\mathbf{r}_i-\mathbf{r}_j)}
    \sum_{m=0}^{\infty} V_{m}L_{m}(q^2),
\end{eqnarray}
where $V_{m}$ is Haldane's pseudopotential parameter
\cite{haldane:1985} and $L_m(x)$ is the Laguerre polynomial. The
momenta $\mathbf{q}$ take discrete values suitable for the unit cell
lattice. $\mathbf{r}_i$ is the guiding center coordinate of the
$i$-th electron.

Our first example of the ideal state at $\nu=1/3$, the Laughlin state $\Psi_{L}$, is obtained as the densest
zero-energy ground state of a short range interaction with only the
pseudopotential $V_1$ nonzero in the above Coulomb Hamiltonian.  Our second
example of the ideal state at half-filled second LL ($\nu=5/2$), the Moore-Read state $\Psi_{Pf}$, is
obtained as the densest zero-energy ground state of a repulsive three-body potential\cite{rezayi:2000}:
\begin{eqnarray}
  H_{3}=-\sum_{i<j<k}S_{i,j,k}
  [\nabla_{i}^4 \nabla_{j}^2 \delta^2(\mathbf{r}_i-\mathbf{r}_j)
  \delta^2(\mathbf{r}_j-\mathbf{r}_k)],
  \label{H3}
\end{eqnarray}
where $S_{i,j,k}$ is a symmetrizer.  We can then compare these ideal
states with the exact ground state of the Coulomb interaction using numerical diagonalization.

\subsection{Computing Indistinguishability in the Fractional Quantum Hall Regime}

Exact diagonalization can be used to compute reduced density
matrices and therefore the indistinguishability in the FQH regime.
In the occupation representation with $N_{\phi}$ orbits, the
$N$-particle FQH states are given by the following general
expression:
\begin{eqnarray}
|\Psi\rangle=\sum_{i}^{N_s}\lambda_{i}|N_i\rangle,
\end{eqnarray}
where $|N_i\rangle=c^{\dagger}_{i_1}\cdot\cdot\cdot
c^{\dagger}_{i_{N}}|0\rangle$ is the $N$-particle basis state with
orbits $i_1,i_2,...,i_N$ occupied and $\lambda_i$ is the normalized
amplitude of the basis state. The operator $c^{\dagger}_{i}$
($c_{i}$) creates (annihilates) a fermion at the $i$-th orbit. $N_s$
is the size of the $N$-particle Hilbert space.

To compute the indistinguishability we must compute the $n$-particle  reduced density matrix, $\rho^{(n)}$. The
total $N$-particle density matrix is given by $\rho_{T}\equiv \vert
\Psi\rangle \langle \Psi \vert$. $\rho^{(n)}$ can be computed using $\text{Tr}_{N-n}(\rho_T) =\sum_{\alpha} \langle \alpha \vert \rho_T \vert \alpha\rangle$, where $\alpha$ denotes all $m$-particle basis states:
$\vert \alpha \rangle = c^{\dagger}_{i_1}\cdot\cdot\cdot
c^{\dagger}_{i_{m}} \vert 0\rangle$ with $m=N-n$.  The
reduced density matrix can now be decomposed in the $n$-particle
basis of $|n_{i}\rangle$ in the occupation representation.  The
reduced density matrix elements are then given by:
\begin{eqnarray}
\rho_{a,b}^{(n)}&=&\langle n_{a}|\text{Tr}_{m}(\rho_{T})|n_{b}\rangle/N_{c} \nonumber \\
&=&\langle n_{a}|\sum_{k}^{m_s}(c_{k_m} \cdot\cdot\cdot c_{k_1})\rho_{T}(c_{k_1}^{\dagger}\cdot\cdot\cdot c_{k_m}^{\dagger})|n_{b}\rangle/N_{c} \nonumber \\
&=&\sum_{i,j}^{N_s} \sum_{k}^{m_s}\lambda_{i} \lambda_{j}^{*} T(a,k,i)T^{*}(b,k,j)/N_{c},\nonumber
\end{eqnarray}
where $ T(a,k,i)=\langle n_{a}|c_{k_m}\cdot\cdot\cdot
c_{k_1}|N_{i}\rangle$,
$N_{c}=(\begin{array}{c}m\\N\end{array})$ is a normalization
constant, and $m_s$ is the size of the Hilbert space for $m$-particle
states.

For a pure system on the torus geometry, $N$-particle states
$|\Psi\rangle$ can be calculated exactly in the momentum subspace.
The momentum operator is given by $J_{N}=\text{Mod}(\sum_{k=1}^{N}
i_{k},N_{\phi})$ for all basis states $|N_i\rangle$. To get nonzero
matrix elements $\rho_{a,b}^{(n)}$, we require that the $n$-particle
states $|n_{a}\rangle$ and $|n_{b}\rangle$ have the same momentum
$J_{n}=\text{Mod}(J_{N}-J_{m},N_{\phi})$ with
$J_{m}=\text{Mod}(\sum_{p=1}^{m} k_{p},N_{\phi})$. Noting this
conservation of momentum rule, we calculate the trace norm term of
Eq.~\ref{In} in momentum sub-blocks.  This use of translational
symmetry considerably reduces the Hilbert space size.

We compare different states using exact diagonalization on the torus
and the above expressions for the indistinguishability. In the
following subsections, we show results with hexagonal unit cells.  We have checked that different choices for unit cells do not impact our conclusions.

\subsection{Distinguishing the Laughlin and Charge Density Wave States}

We first compare the uniform Laughlin state with a gapless state at
one-third filling. A transition between the Laughlin state and the
gapless state can be driven by softening the short range part of the
Coulomb interaction, $V_{1}\rightarrow V_{1}-dV_{1}$.  This gapless
state is a non-topological CDW.  Such CDW states have been discussed
in the literature
\cite{laughlin2gapless,koulakov:1996,lee:2001,shibata:2004} at a
variety of fillings.

\begin{figure}[t]
\centerline{\includegraphics [width=3.2 in] {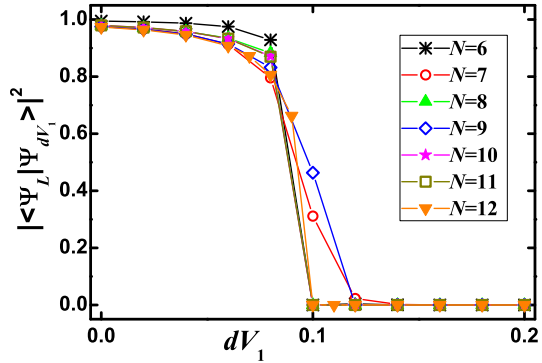}} \caption{(Color
online) Squared wavefunction overlap between Laughlin state and the
ground state of $\nu=1/3$ system with modulated pseudopotentials $V_1$.}
\label{overlap}
\end{figure}

Figure~\ref{overlap} shows the overlap between the Laughlin and the
lowest energy state, $\Psi_{dV_1}$, from the softened Coulomb model as a function of $dV_{1}$
for several system sizes.  Here we see that in the lowest LL the Coulomb
point ($dV_{1}=0$) lies squarely in the Laughlin liquid regime.  But
as the short ranged part of the Coulomb interaction is softened, the
Laughlin gap collapses (not shown) to reveal a transition towards
the CDW phase. The CDW phase is non-uniform and may occur at a momentum different from
the Laughlin state. We note that there are many nearly
degenerate states in the CDW regime.  There is a small energy
splitting in finite sized systems. We take the lowest energy state.

We now explore the nature of the liquid-to-CDW transition using
the indistinguishability.  The Laughlin state is a topological state
with degenerate topological sectors but the CDW state does not
represent a topologically ordered state.  It is best described by local correlators.  We therefore expect a
distinct signature in $I_{n}$ in the transition.

\begin{figure}
\centerline{\includegraphics [width=3.2 in] {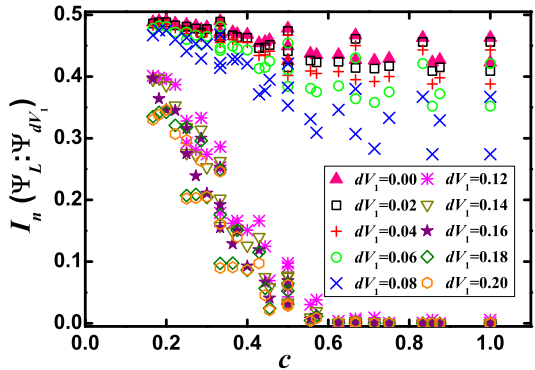}} \caption{(Color
online) Indistinguishability between Laughlin state and calculated
ground state as a function of $c$ for $\nu=1/3$ system with different
pseudopotentials. The data for each $dV_1$ are from different system sizes $N=6,7,8,9,10,11$, and $12$.} \label{bubble}
\end{figure}

Figure~\ref{bubble} shows $I_{n}$ versus $c$ for several different
pseudopotentials as we cross the transition from the Laughlin liquid
(top points) to the CDW state (bottom points).  The data for each
$dV_{1}$ represent composites from several different values of $N$
indicating that even for small $N$ we have approximate data collapse
for $I_{n}$ (away from the transition point).  Near the transition
point ($dV_{1}\sim 0.10$) the data scatter.  From the figure we see
that above the transition ($0\leq dV_{1}\lesssim 0.08$) the Laughlin
state is barely distinguishable from the ground state of the softened Coulomb
interaction.  Once we cross the transition ($dV_{1}\gtrsim 0.12$)
the Laughlin state can be distinguished from the CDW state but
\emph{only} with measurements on $n\sim N$ particles.  This is
surprising because the CDW state is locally non-uniform and one
would expect it to be locally different from the
uniform Laughlin state.  We note however that here we have compared
only the lowest energy CDW state that arises in our finite size
calculation.  Inclusion of all low energy CDW states that arise
in the thermodynamic limit may lower the $n=\mathcal{O}(N)$
dependence to an $n=\mathcal{O}(1)$ dependence.

\subsection{Distinguishing Topological Sectors in the Fractional Quantum Hall Regime}

We now examine the indistinguishability between two distinct
topological sectors.  We first compare two Laughlin states. On the
torus there are three degenerate Laughlin states. These orthogonal
Laughlin states define distinct topological sectors. We now ask if
two distinct sectors can be distinguished with
$n=\mathcal{O}(\sqrt{N})$ correlation functions.

\begin{figure}[t]
\centerline{\includegraphics [width=3.2 in] {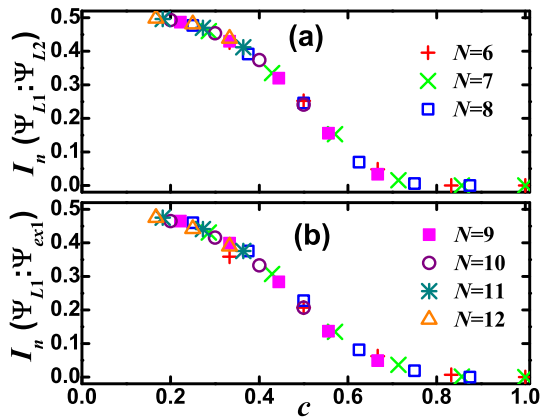}} \caption{(Color
online) Indistinguishability as a function of $c$ for $\nu=1/3$
system (a) between two degenerate Laughlin states, and (b) between
Laughlin state and the first excited state.} \label{laughlin}
\end{figure}

In Figure~\ref{laughlin}a we plot the indistinguishability computed
for two degenerate Laughlin states, $\Psi_{L1}$ and $\Psi_{L2}$. The
approximate data collapse has been shown for different system sizes,
indicating that the results are valid for the thermodynamic limit.
We find that the two states are nearly indistinguishable for small
$c$ but become distinguishable only for large $c\gtrsim 0.5$. We do
not find an $n=\mathcal{O}(\sqrt{N})$ dependence. Instead, the
non-local, $\mathcal{O}(N)$, distinction between states is found to
be a generic feature of \emph{any} two quantum Hall states.
Figure~\ref{laughlin}b compares the Laughlin state $\Psi_{L1}$ with
it's first excited state $\Psi_{ex1}$ at zero momentum. The first
excited state can be thought of as a composite fermion particle-hole
pair that is formed from superpositions of all electron coordinates
\cite{jain:1989,jain:2007}.  We find precisely the same $I_{n}$
dependence here indicating that the structure we observe is generic
for any two orthogonal lowest LL states derived from short range
models.  For long range interactions we find that the $I_{n}$ versus
$n$ dependence exhibits the similar behavior.

\begin{figure}[t]
\centerline{\includegraphics [width=3.2 in] {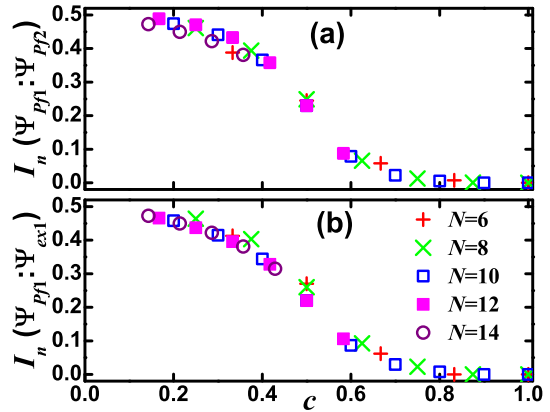}} \caption{(Color
online) Indistinguishability as a function of $c$ for the $\nu=5/2$
system (a) between two degenerate Moore-Read states, and (b) between
the Moore-Read ground state and the first excited state.} \label{mooreread}
\end{figure}

We have also checked the non-Abelian Moore-Read states. From
Eq.~\ref{H3} we can generate six degenerate Moore-Read states on the
torus, corresponding to six distinct topological sectors. We plot
the indistinguishability between two distinct Moor-Read states,
$\Psi_{Pf1}$ and $\Psi_{Pf2}$, in Figure~\ref{mooreread}a and the
indistinguishability between $\Psi_{Pf1}$ and its first excited
state, $\Psi_{ex1}$, in Figure~\ref{mooreread}b for different system
sizes. These figures show precisely the same generic structure as
Abelian Laughlin states.

\section{Summary}
\label{summary}

We have computed the indistinguishability, $I_{n}$, (Eq.~\ref{In})
between distinct topological states in two different types of
two-dimensional models, the toric code lattice model of spins and
FQH models of 2DEGs in a strong magnetic field.  Both models show
ground states with topological degeneracies. Using $I_{n}$ we were
able to show that the nature of the topological order in the toric
code is distinct from that in the FQH regime.

In the toric code, basis states are localized in two dimensions to
lie at discrete sites. The model was constructed to obey strictly
one-dimensional symmetries. These symmetries then yield topological
degeneracies when the model is placed on a surface with periodic
boundaries.  As a result our calculation of  $I_{n}$ showed that
measurements on a carefully chosen  set of $n\sim \sqrt{N}$ spins in
an $N$-particle system can accurately distinguish topological
sectors, as expected. The topological quantum Hall states, in
contrast, always show an $n\sim N$ dependence.  This is a result of
the one-dimensional nature of the Hilbert space itself.  FQH
correlators distinguishing two states \cite{haldane:1983} span all
particles even though the correlators are constructed from a
one-dimensional product of operators.

Our study used the indistinguishability to show that quantum Hall
states in periodic systems can only be distinguished with
correlators of the order of the system size, $N$.  This implies that
the wavefunction is sufficient \emph{and} necessary in a full
description of FQH states.  Our results also imply that in comparing
candidate quantum Hall states in systems without edges, overlap is an efficient tool to distinguish two
states (overlap is equivalent to $I_{n=N}$ up to a constant factor).  Our results show that small-size correlation functions ($n\lesssim N/2$) can not accurately distinguish two quantum Hall
states in the torus geometry.

It would be interesting to extend our analysis to systems with
edges.  Correlators of FQH edge states can be used
to distinguish states in the bulk \cite{wen:1990b}.  A
calculation of $I_{n}$ for systems with edges may show a different
$n$ dependence for states chosen near the system edge.

VWS acknowledges support from the Jeffress Memorial Trust, Grant No.
J-992.

\end{document}